\documentclass[
aps,
prc,
showpacs,
twocolumn,
superscriptaddress,
nofootinbib,
floatfix]
{revtex4-1}
\usepackage{graphicx,
longtable,
slashed}%

\begin{document}

\title{Neutrinoless double-beta decay and seesaw mechanism}

\author{Samoil M. Bilenky}
\affiliation{Physik-Department E15, Technische Universit\"at M\"unchen,
D-85748 Garching, Germany}
\affiliation{Laboratory of Theoretical Physics, JINR,
141980 Dubna, Moscow region, Russia}
\author{        Amand Faessler}
\affiliation{Institute of Theoretical Physics,
University of Tuebingen, 72076 Tuebingen, Germany}
\author{Walter Potzel}
\affiliation{Physik-Department E15, Technische Universit\"at M\"unchen,
D-85748 Garching, Germany}
\author{Fedor \v Simkovic}
\affiliation{Laboratory of Theoretical Physics, JINR,
141980 Dubna, Moscow region, Russia}
\affiliation{Department of Nuclear Physics and Biophysics,
Comenius University, Mlynsk\'a dolina F1, SK--842 15
Bratislava, Slovakia}

\date{\today}

\begin{abstract}
From the standard seesaw mechanism of neutrino mass generation,
which is based on the assumption that the lepton number is violated
at a large  ($\sim 10^{15}$ GeV) scale, follows that
the neutrinoless double-beta decay ($0\nu\beta\beta$-decay) is
ruled by the Majorana neutrino mass mechanism.
Within this notion, for the inverted neutrino-mass hierarchy we derive
allowed ranges of half-lives of the $0\nu\beta\beta$-decay for nuclei
of experimental interest with different sets of nuclear matrix elements.
The present-day results of the
calculation of the $0\nu\beta\beta$-decay nuclear matrix elements
are briefly discussed. We argue that if $0\nu\beta\beta$-decay will
be observed in future  experiments sensitive to the effective Majorana mass 
in the inverted mass hierarchy region, a comparison of the derived ranges 
with measured  half-lives will allow us to probe the standard seesaw mechanism 
assuming that future cosmological data will
establish the sum of neutrino masses to be about 0.2 eV.
\end{abstract}


\nopagebreak[4]


\maketitle

\section{Introduction \label{SecI}}

The observation of neutrino oscillations in atmospheric \cite{SK},
solar \cite{Solar}, reactor \cite{Kamland} and accelerator \cite{K2K,Minos}
neutrino experiments is the most important recent discovery in particle
physics. It is very unlikely that neutrino masses, many orders of magnitude
smaller than the masses of quarks and leptons, are generated by the standard
Higgs mechanism. Small neutrino masses and neutrino mixing are commonly
considered as  a signature of physics beyond the Standard Model (SM).
Several beyond the SM mechanisms of neutrino mass generation  were proposed.
The most viable and plausible mechanism is the famous seesaw mechanism which
is based on the assumption that the total lepton number $L$ is violated at
a scale much larger than the electroweak scale
$v=(\sqrt{2}G_{F})^{-1/2}\simeq 246$ GeV.

If the total lepton number is violated, neutrinos $\nu_i$ with definite
masses are Majorana particles. After the discovery of neutrino oscillations
the problem of the nature of neutrinos with definite masses
(Majorana or Dirac?) became the most pressing issue.

Information about the nature of neutrinos with definite masses can not be
obtained via the investigation of neutrino oscillations \cite{BilHos}. In
order to obtain such an information it is necessary to study processes in
which the total lepton number $L$ is violated. The investigation of
neutrinoless double-beta decay ($0\nu\beta\beta$-decay) of even-even nuclei
$$(A,Z)\to (A,Z+2)+e^{-}+e^{-}$$
is the most sensitive way to search for the effects of the lepton number violation.

The observation of the $0\nu\beta\beta$-decay will prove that $\nu_{i}$ are
Majorana particles. In this paper we show that an evidence of this process in
future $0\nu\beta\beta$-decay experiments sensitive to the effective Majorana mass 
in the inverted mass hierarchy region could allow us to obtain information about 
the validity  of the original seesaw idea \cite{seesaw} of
neutrino mass generation associated with a violation of the total lepton number
at GUT scale  assuming that a proper information about the lightest neutrino mass 
will be available from future cosmological data.

\section{Seesaw mechanism of neutrino mass generation}

The standard seesaw mechanism (type I seesaw)
is based on the assumption that there exist heavy Majorana leptons $N_{i}$,
singlets of the $SU_{L}(2)\times U(1)$ group, which have the following lepton
number violating Yukawa interactions with lepton and Higgs doublets
\begin{equation}\label{heavyH}
\mathcal{L}=-\sqrt{2}\sum_{i,l}Y_{li}\overline  L_{lL} N_{iR} \tilde{H }+
\mathrm{h.c.}.
\end{equation}
Here
\begin{eqnarray}\label{heavyH1}
L_{lL}
=
\left(
\begin{array}{c}
\nu_{lL}
\\
l_L
\end{array}
\right),
\qquad
H
=
\left(
\begin{array}{c}
H^{(+)}
\\
H^{(0)}
\end{array}
\right)
\end{eqnarray}
are lepton and Higgs doublets, $\tilde{H}=i\tau_{2}H^{*}$, $Y_{il}$ are
dimensionless constants and $$N_{i}=N^{c}_{i}=C\bar N^{T}_{i} $$
is the field of heavy Majorana leptons with mass $M_{i}$ which is much
larger than $v$.

At electroweak energies for the processes with virtual  $N_{i} $
the interactions (\ref {heavyH}) generate the effective Lagrangian
 \begin{equation}\label{heavyH2}
\mathcal{L}_{\rm{eff}}=-\frac{1}{\Lambda}
\sum_{l',l,i}\,\overline L_{l'L}\tilde{H }\sum_{i}(Y_{l'i}\frac{\Lambda}{M_{i}}Y_{li})C
\tilde{H }^{T}(\overline L_{lL})^{T} +
\mathrm{h.c.},
\end{equation}
which does not conserve the total lepton number $L$ and is the only
effective Lagrangian of the dimension five \cite{Weinberg}. In (\ref {heavyH2}),
the parameter $\Lambda$ has the dimension of mass and characterizes
the scale of new physics beyond the SM.

After spontaneous violation of the electroweak symmetry the Lagrangian
(\ref{heavyH2}) generates {\em the left-handed Majorana  mass term}
\begin{equation}\label{heavyH4}
\mathcal{L}^{\mathrm{M}}=-
\frac{1}{2}
\,\sum_{l',l}
\overline \nu_{l'L}~
M^{L}_{l'l}~
(\nu_{lL})^{c}
+
\mathrm{h.c.}=-\frac{1}{2}
\,\sum_{i}m_{i}\bar \nu_{i} \nu_{i},
\end{equation}
where
\begin{equation}\label{Majmass}
M^{L}=Y\frac{v^{2}}{M}Y^{T}=UmU^{T},
\end{equation}
$\nu^{c}_{i}=\nu_{i}$ is the field of the Majorana neutrino with the mass
$m_{i}$ and the flavor field $\nu_{lL}$ is given by the standard mixing relation
\begin{equation}\label{Majmass1}
\nu_{lL} = \sum_{i} U_{li} \nu_{iL}.
\end{equation}
Here,  $U_{li}$ are the elements of the Pontecorvo-Maki-Nakagawa-Sakata
neutrino mixing matrix \cite{BPont,MNS}.
The size of neutrino masses is determined by the seesaw factor $\frac{v^{2}}{M_{i}}$.
From the existing data we can estimate that  $M_{i}\simeq (10^{14}- 10^{15})$ GeV.

Let us stress that from the point of view of the standard seesaw approach
{\em small Majorana neutrino masses  are the only low energy signature of
physics beyond the SM at a GUT scale where the total lepton number is violated.}\footnote{In the early Universe at very high temperatures heavy Majorana leptons $N_{i}$
can be produced. Their $CP$-violating decays could lead to the baryon asymmetry
of the Universe (see \cite{Nir} and references therein).}

The effective Lagrangian $\mathcal{L}_{\rm{eff}}$
and, consequently, the left-handed Majorana mass term (\ref {heavyH4}) can be
generated not only by the interaction (\ref{heavyH}) but also by an interaction of
lepton pairs and a Higgs pair with a triplet heavy scalar boson $\Delta$
(type II seesaw) and by an interaction of lepton-Higgs  pairs with heavy Majorana
triplet fermion $\Sigma$ (type III seesaw) (see \cite{Ma}).

From the previous discussion we can conclude that if the total lepton number $L$ is violated at a GUT scale due to the existence of a heavy singlet (or triplet) Majorana fermion or a heavy triplet scalar boson interacting with standard lepton and Higgs doublets then
\begin{itemize}
  \item neutrinos have small, seesaw suppressed masses (in accordance with the existing experimental data),
  \item neutrinos with definite masses are Majorana particles and the only mechanism of the $0\nu\beta\beta$-decay is the exchange of virtual Majorana neutrinos.
\end{itemize}
In this paper we will explore these just mentioned general consequence of the standard seesaw mechanism.

\section{$0\nu\beta\beta$-decay: Nuclear matrix elements}

\begin{table*}[!t]
\caption{The NME of the $0\nu\beta\beta$-decay calculated in the framework of
different approaches:
interacting shell model (ISM) \protect\cite{lssm}, quasiparticle random
phase approximation (QRPA) \protect\cite{qrpa1,qrpa2}, projected Hartree-Fock Bogoliubov
approach (PHFB, PQQ2 parametrization) ) \protect\cite{phfb},
energy density functional method (EDF) \cite{edf} and
interacting boson model (IBM) \protect\cite{ibm}. The Miller-Spencer
Jastrow two-nucleon short-range correlations are taken
into account.  The EDF results are multiplied by 0.80 in order to account for the difference
between UCOM and Jastrow \protect\cite{anatomy}.
$g_A = 1.25$ and $R = 1.2 A^{1/3}$ are assumed. }
\label{tab.1}
\begin{tabular}{lccccc}\hline\hline
Transition  & \multicolumn{5}{c}{$M(A,Z)$} \\ \cline{2-6}
            & ISM \cite{lssm} & QRPA \cite{qrpa1,qrpa2} & IBM-2 \cite{ibm}
            & PHFB \cite{phfb} & EDF \cite{edf}  \\  \hline
${^{48}Ca}  \rightarrow {^{48}Ti}$ & 0.61  &        &      &       &  1.91 \\
${^{76}Ge}  \rightarrow {^{72}Se}$ & 2.30  &  4.92  & 5.47 &       &  3.70 \\
${^{82}Se}  \rightarrow {^{82}Kr}$ & 2.18  &  4.39  & 4.41 &       &  3.39 \\
${^{96}Zr}  \rightarrow {^{96}Mo}$ &       &  1.22  &      &  2.78 &  4.54 \\
${^{100}Mo} \rightarrow {^{100}Ru}$ &      &  3.64  & 3.73 &  6.55 &  4.08 \\
${^{116}Cd} \rightarrow {^{116}Sn}$ &      &  2.99  &      &       &  3.80 \\
${^{124}Sn} \rightarrow {^{124}Te}$ & 2.10 &        &      &       &  3.87 \\
${^{128}Te} \rightarrow {^{128}Xe}$ & 2.34 & 3.97   & 4.52 &  3.89 &  3.30 \\
${^{130}Te} \rightarrow {^{130}Xe}$ & 2.12 & 3.56   & 4.06 &  4.36 &  4.12 \\
${^{136}Xe} \rightarrow {^{136}Ba}$ & 1.76 & 2.30   &      &       &  3.38 \\
${^{150}Nd} \rightarrow {^{150}Sm}$ &      & 3.16   & 2.32 &  3.16 &  1.37 \\
\hline\hline
\end{tabular}
\end{table*}

Neutrinoless double $\beta$-decay of even-even nuclei is a process of
second order in the Fermi constant $G_{F}$ with the exchange of virtual
Majorana neutrinos between $n-p-e^{-}$ vertexes. The mixed neutrino
propagator has the form
\begin{eqnarray}\label{propagator}
&&\sum_{i} U^{2}_{ei}\left (\frac{1-\gamma_{5}}{2}\right )
\frac{\gamma\cdot p+m_{i}}{p^{2}-m^{2}_{i}}
\left (\frac{1-\gamma_{5}}{2}\right ) C\nonumber\\
&&\simeq m_{\beta\beta}
\frac{1}{p^{2}}\left (\frac{1-\gamma_{5}}{2}\right ) C,
\end{eqnarray}
where
\begin{equation}\label{MJmass}
m_{\beta\beta}=\sum_{i} U^{2}_{ei}m_{i}
\end{equation}
is the effective Majorana mass.

Let us stress that
\begin{itemize}
\item due to the $V-A$ structure of the weak charged current and
neutrino mixing the matrix elements of the $0\nu\beta\beta$-decay
are proportional to $m_{\beta\beta}$.
\item the average momentum of the virtual neutrinos is about 100 MeV.
Thus,  $p^{2}\gg m^{2}_{i}$ and the nuclear matrix elements do not
depend on $m_{i}$. As a result, in the matrix elements of the
$0\nu\beta\beta$-decay, neutrino properties and nuclear properties
are factorized.
\end{itemize}
The inverted half-life of the $0\nu\beta\beta$-decay is given by
the following general expression \cite{Doi}
\begin{equation}\label{halflife}
\frac{1}{T^{0\,\nu}_{1/2}(A,Z)}=
|m_{\beta\beta}|^{2}\,|M(A,Z)|^{2}\,G^{0\,\nu}(E_{0},Z).
\end{equation}
Here $M(A,Z) $ is the nuclear matrix element (NME) (matrix element
between states of the initial and the final nuclei of the integrated
product of two hadron charged currents and the neutrino propagator)
and  $G^{0\,\nu}(E_{0},Z)$ is a known phase-space factor ($E_{0}$ is
the energy release).\footnote{For numerical values of
$G^{0\,\nu}(E_{0},Z)$ see\cite{kinematics}.}

For our calculations we will need the values of the $0\nu\beta\beta$-decay
NMEs for different nuclei of experimental interest. We will briefly discuss here
the present-day situation of the calculation of NMEs, compare existing methods,
stress differences between them and present current values of NMEs.

The calculation of the NME is a complicated nuclear many-body problem.
During many years two  approaches were used: the Quasiparticle Random
Phase Approximation (QRPA)\cite{qrpa1,anatomy,qrpa2} and the Interacting
Shell Model (ISM)\cite{lssm}. There are substantial
differences between both approaches. The QRPA treats a large single
particle model space, but truncates heavily the included configurations.
The ISM, by contrast, treats a small fraction of this model space,
but allows the nucleons to correlate in many different ways.
We note that the latest QRPA results of the Jyv\"askyl\"a-La Plata
group \cite{suhonen} agree well with those of
the Tuebingen-Bratislava-Caltech group \cite{qrpa1,anatomy}
by using the same way of adjusting the parameters of the nuclear
Hamiltonian  \cite{qrpa1}.

In the last few years several new approaches have been used for
the calculation of the $0\nu\beta\beta$-decay NMEs: the angular
momentum Projected Hartree-Fock-Bogoliubov method (PHFB)
\cite{phfb}, the Interacting Boson Model (IBM) \cite{ibm},
and the Energy Density Functional method (EDF) \cite{edf}.
In the PHFB approach, the nucleon pairs different from $0^+$ in the
intrinsic coordinate system are strongly suppressed. In the framework
of the ISM and the QRPA  approaches it was shown, however, that other neutron pairs
make significant contributions \cite{seniority}. Let us notice also that
in the IBM approach only transitions of $0^+$ and $2^+$ neutron pairs
into proton pairs  are taken into account. The EDF approach is
an improvement with respect to the PHFB approach. Beyond-mean-field
effects are included within the generating coordinate method with
particle number and angular momentum projection for both initial
and final ground states. But, the quality of the IBM and
the EDF many-body wave functions have not been tested yet
by the calculation of the $2\nu\beta\beta$-decay half-lives.

In Table \ref{tab.1}, recent results of the different methods are summarized.
The presented numbers have been obtained with
the unquenched value of the axial coupling constant ($g_A=1.25$)\footnote{
A modern value of the axial-vector coupling  constant is $g_A =1.269$. We note
that in the referred calculations of the $0\nu\beta\beta$-decay NMEs
the previously accepted value $g_A=1.25$ was assumed.},
Miller-Spencer Jastrow short-range
correlations \cite{miler} (the EDF values are multiplied by 0.80 in order to account
for the difference between the unitary correlation operator method
(UCOM) and the Jastrow approach \protect\cite{anatomy}),
the same nucleon dipole form-factors, higher order corrections to
the nucleon current and the nuclear radius $R = r_0 A^{1/3}$, with
$r_0 = 1.2$ fm (the QRPA values \cite{qrpa1}
for $r_0 =1.1$ fm are rescaled with the factor $1.2/1.1$). Thus,
the discrepancies among the results of different
approaches are solely related to the approximations on which a given
nuclear many-body method is based.

From Table  \ref{tab.1} we see that the smallest values of NMEs
are obtained in the ISM approach. They are by about a factor of 2-3
smaller in comparison with results of other methods.
The largest values of NME  are obtained
in the IBM (${^{76}Ge}$ and ${^{128}Te}$), PHFB (${^{100}Mo}$,
${^{130}Te}$ and ${^{150}Nd}$),
QRPA (${^{150}Nd}$) and EDF (${^{48}Ca}$, ${^{96}Zr}$, ${^{116}Cd}$,
${^{124}Sn}$  and ${^{136}Xe}$) approaches.
NMEs obtained by the QRPA and IBM methods are in a good
agreement (with the exception of ${^{150}Nd}$). It is
remarkable that for ${^{130}Te}$ the results of four different methods
(QRPA, PHFB, IBM and EDF) are close to each other.

The differences among the listed methods of NME calculations for
the  $0\nu\beta\beta$-decay are  due to the following reasons:\\
(i) The mean field is used in different ways. As a result,
single particle occupancies of individual orbits of various
methods differ significantly from each other \cite{occup}. \\
(ii) The residual interactions are of various origin and
renormalized in different ways.\\
(iii) Various sizes of the model space are taken into account.\\
(iv) Different many-body approximations are used in the
diagonalization of the nuclear Hamiltonian.\\
Each of the applied methods has some advantages and disadvantages.

\section{Possible probe of the Majorana neutrino mass mechanism of the $0\nu\beta\beta$-decay}

Many experiments on the search for $0\nu\beta\beta$-beta decay
of different nuclei were  performed (see \cite{reviews}). No indications
in favor of $0\nu\beta\beta$-decay were obtained in these experiments.
There exist, however, a claim of the observation of the $0\nu\beta\beta$-decay
of $^{76}\rm{Ge}$ made by some participants of the Heidelberg-Moscow collaboration
\cite{Klapdor}. Their estimated value of the effective Majorana mass 
(assuming a specific value for the NME) is $|m_{\beta\beta}|\simeq 0.4$ eV. This result will be checked
by an independent experiment relatively soon.  In the new germanium
experiment GERDA \cite{Gerda}, the Heidelberg-Moscow sensitivity  will be reached
in about one year of measuring time.

\begin{figure}[!t]
\vspace*{1.0cm}
\hspace*{0.cm}
\includegraphics[scale=.32]{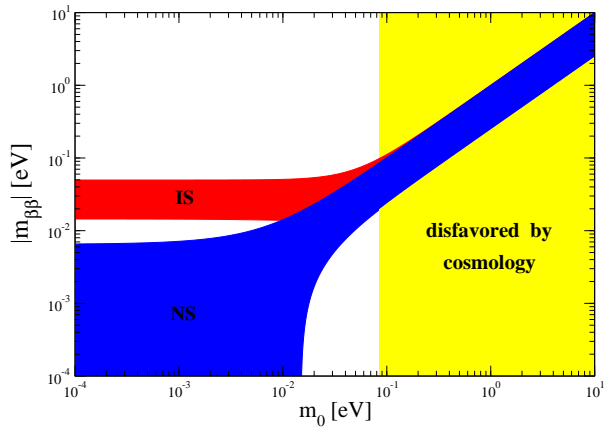}
\caption{(Color online)
Effective Majorana neutrino mass $|m_{\beta\beta}|$ as function of the lightest
neutrino mass $m_0$ for the cases of normal (NS, $m_0=m_1$) and inverted
(IS, $m_0=m_3$) spectrum of neutrino masses.
$\Delta m^2_{A} = (2.43 \pm 0.13) \times 10^{-3}~\mathrm{eV}^{2}$ \cite{Minos},
$\Delta m^2_{S}= (7.65^{+0.13}_{-0.20}) \times 10^{-5}~\mathrm{eV}^{2}$ \cite{global},
$\rm{\tan}^2{\theta_{12}}=0.452_{-0.033}^{+0.035}$ \cite{Kamland} and
$0.03(0.04) < \rm{\sin}^2 2\theta_{13} < 0.28(0.34)$ \cite{T2K} for NS (IS) are considered.
The current limit of
$\sum_{i=1}^3 m_i \le 0.28$ eV  \cite{Thomas} for the sum of neutrino masses
excludes values of $m_0$ larger 0.084 eV.
\label{fig.1}}
\end{figure}

From the most precise  experiments on the search for  $0\nu\beta\beta$-decay the
following bounds were inferred \protect\cite{baudis, cuore, nemo}:
\begin{eqnarray}\label{bounds}
|m_{\beta\beta}| & < & (0.20-0.32) ~eV~~ (^{76}\mathrm{Ge}), \nonumber\\
               & < & (0.30-0.71) ~eV~~ (^{130}\mathrm{Te}), \nonumber\\
               & < & (0.50-0.96) ~eV~~ (^{130}\mathrm{Mo}).
\end{eqnarray}
These bounds we obtained using the $0\nu\beta\beta$-decay NMEs of \cite{src} calculated with
Brueckner two-nucleon short-range correlations.

In future experiments, CUORE\cite{cuore}, EXO\cite{Exo}, MAJORANA\cite{Majorana},
SuperNEMO \cite{supernemo}, SNO+ \cite{snoplus},  Kamland-ZEN
and others \cite{reviews}, a sensitivity
\begin{equation}\label{sensitiv}
|m_{\beta\beta}|\simeq \mathrm{a~few}~10^{-2}~\mathrm{eV}
\end{equation}
is planned to be reached.

The value of the effective Majorana mass strongly depends on the character
of the neutrino mass spectrum. For the case of three neutrinos two types
of  mass spectra are allowed by the neutrino oscillation data:
\begin{enumerate}
  \item Normal spectrum (NS)
 \begin{equation}\label{Nspectrum}
 m_{1}<m_{2}<m_{3},\quad \Delta m^{2}_{12}\ll\Delta m^{2}_{23},
 \end{equation}
 \item Inverted spectrum (IS)
 \begin{equation}\label{Ispectrum}
 m_{3}<m_{1}<m_{2},\quad \Delta m^{2}_{12}\ll|\Delta m^{2}_{13}|.
 \end{equation}
\end{enumerate}
In the case of NS, we have for the neutrino masses
\begin{equation}\label{numassN}
m_{2}=\sqrt{m^{2}_{0}+\Delta m^{2}_{S}},\quad
 m_{3}\simeq \sqrt{m^{2}_{0}+\Delta m^{2}_{A}}.
\end{equation}
For IS we find
\begin{equation}\label{numassI}
m_{1}=\sqrt{m^{2}_{0}+\Delta m^{2}_{A}},\quad
 m_{2}\simeq \sqrt{m^{2}_{0}+\Delta m^{2}_{A}}.
\end{equation}
Here $\Delta m^{2}_{12}=\Delta m^{2}_{S}$ and $\Delta m^{2}_{23}(
|\Delta m^{2}_{13}|)=\Delta m^{2}_{A}$ are the solar and atmospheric neutrino mass-squared differences, respectively, and $m_{0}=m_{1}(m_{3})$ is the lightest neutrino mass for NS(IS).

In  Fig. \ref{fig.1} the effective Majorana mass $|m_{\beta\beta}|$
is plotted as a function of $m_{0}$ for the cases of the NS and the IS\footnote{ Notice that for NS this figure differs significantly from analogous 
figures published in the literature. This is connected with the fact that we use  
new T2K data \cite{T2K} for the values of the parameter $\sin^{2}2\theta_{13}$.}.
The lowest value for the sum of the neutrino masses, which can be
reached in future cosmological measurements \cite{Thomas,Serpico,Abazajian},
is about (0.05-0.1) eV. The corresponding values of $m_0$ are in the region,
where the IS and the NS predictions for $|m_{\beta\beta}|$ differ significantly from each other.

Future experiments on the search for the $0\nu\beta\beta$-decay
will probe the region of the inverted mass hierarchy\footnote{Let us note that
the $0\nu\beta\beta$-decay in the case of the inverted hierarchy was considered
in detail in a recent paper \cite{Dueck}.}
\begin{equation}\label{inverted}
m_{3}\ll m_{1}<m_{2}.
\end{equation}
In this case we have
\begin{equation}\label{inverted1}
m_{1}\simeq m_{2}\simeq \sqrt{\Delta m^{2}_{A}},\quad
m_{3}\ll \sqrt{\Delta m^{2}_{A}}.
\end{equation}
Neglecting small contribution of the term $m_{3}|U_{e3}|^{2}$,
for the effective Majorana mass in the case of the inverted mass hierarchy (\ref{inverted}) we obtain the following expression
\begin{equation}\label{invhierarA}
|m_{\beta\beta}|\simeq \sqrt{ \Delta m^{2}_{A}}\,~ \cos^{2}\theta_{13}~
(1-\sin^{2} 2\,\theta_{12}\,\sin^{2}\alpha_{12})^{\frac{1}{2}},
\end{equation}
where $\alpha_{12}=\alpha_{2}-\alpha_{1}$ is the difference
of the Majorana phases of the elements $U_{e2} $ and $U_{e1} $ with $U_{ei}=|U_{e1}|e^{i\alpha_{i}}$  (i=1,2).

The phase
difference $\alpha_{12}$ is the only unknown parameter in the expression
for $|m_{\beta\beta}|$. From (\ref{invhierarA}) we obtain the following inequality
\begin{equation}\label{invhierar}
\sqrt{ \Delta m^{2}_{A}} \cos^{2}\theta_{13}~ \cos2\,\theta_{12}\,    \leq |m_{\beta\beta}|\leq  
\sqrt{ \Delta m^{2}_{A}}\cos^{2}\theta_{13},
\end{equation}
where upper and lower bounds correspond to the case of the $CP$ invariance
in the lepton sector (upper (lower) bound corresponds to the same (opposite)
$CP$-parities of $\nu_{1}$ and $\nu_{2}$).

From (\ref{invhierar}) we find
\begin{equation}\label{invhierar1}
 1.5\cdot10^{-2} ~\mathrm{eV}   \leq |m_{\beta\beta}|\leq 5.0\cdot 10^{-2}~\mathrm{eV},
\end{equation}
where we used the MINOS value  $\Delta m^2_{A}=
(2.43 \pm 0.13) \times 10^{-3}~\mathrm{eV}^{2}$ \cite{Minos},
the solar-KamLAND value $\tan^2{\theta_{12}}=0.452_{-0.033}^{+0.035}$ \cite{Kamland} 
and the recent T2K observation $\theta_{13}$:   $0.04 < \rm{\sin}^2 2\theta_{13} < 0.34$ \cite{T2K}.

From (\ref{inverted1}) follows that in the case of inverted mass hierarchy we have for the sum of the neutrino masses
\begin{equation}\label{sum}
    \sum^{3}_{i=1}m_{i}\simeq 2\sqrt{ \Delta m^{2}_{A}}\simeq 10^{-1}~\mathrm{eV}.
\end{equation}
As is well known, the quantity  $\sum_{i}m_{i}$ can be inferred from
the measurement of the distribution of galaxies  and other cosmological observations. At present from cosmological data the bound $\sum_{i}m_{i}\lesssim 0.5$ eV was obtained \cite{Hannestad},\cite{Thomas}. It is expected that in the future various cosmological observables will be sensitive to $\sum_{i}m_{i}$ in the range $(6\cdot 10^{-3}-10^{-1})$ eV (see, for example, \cite{Abazajian}). Thus, the inverted neutrino mass hierarchy (\ref{inverted}) will be tested by future precision cosmology.

We will now discuss a possibility to check the Majorana mass mechanism for the case that the $0\nu\beta\beta$-decay will be observed in
future experiments sensitive to the region (\ref{invhierar1}) of the inverted hierarchy.

\begin{figure}[!t]
\vspace*{1.0cm}
\hspace*{0.cm}
\includegraphics[scale=.32]{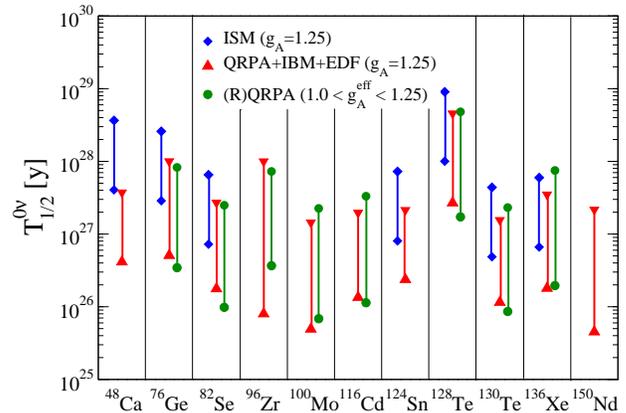}
\caption{(Color online)
The allowed intervals of the $0\nu\beta\beta$-decay half-life  $T^{0\nu}_{1/2}$
[see Eq. (\ref{final})]
for nuclei of experimental interest in the case
of the see-saw mechanism and inverted hierarchy of neutrino masses.
Results are presented for the ISM and a group of 3 methods
(QRPA, IBM and EDF) by considering
$g_A=1.25$ and Jastrow short-range correlations. In addition the
allowed (R)QRPA ranges of NMEs are considered by assuming
$1.0 \le g^{eff}_A \le 1.25$ and the Brueckner short-range correlations
\protect\cite{src}.
\label{fig.2}}
\end{figure}

From (\ref{halflife}) and (\ref{invhierar})
we find the following inequalities for the half-life of $0\nu\beta\beta$-decay
\begin{equation}\label{final}
T^{min}_{1/2}(A,Z)
\leq T^{0\,\nu}_{1/2}(A,Z) \leq
T^{max}_{1/2}(A,Z)
\end{equation}
with
\begin{eqnarray}
T^{min}_{1/2}(A,Z) &=&
 \frac{1}{\Delta m^{2}_{A}|M(A,Z)|^{2}G^{0\,\nu}(E_{0},Z)},
\nonumber\\
T^{max}_{1/2}(A,Z) &=&
\frac{1}{\Delta m^{2}_{A} \cos^{2}2\,\theta_{12}
|M(A,Z)|^{2}G^{0\,\nu}(E_{0},Z) }.\nonumber\\
\end{eqnarray}
In Fig.\ref{fig.2} we present ranges of $0\nu\beta\beta$-decay
half-lives of different nuclei.
If the measured half-life of the $0\nu\beta\beta$-decay is in  the range
given by Eq.  (\ref{final}) this will be an evidence in favor
of the Majorana neutrino mass mechanism (assuming inverted mass hierarchy).

Two remarks are in order:
\begin{enumerate}
\item
It is seen from Fig.\ref{fig.1} that the horizontal band
determined by the inequality (\ref{invhierar1}) is restricted by the condition $m_{0}\le \sqrt{\Delta m^{2}_{A}}\simeq 5\cdot 10^{-2}$ eV (which corresponds to $\sum_{i}m_{i}\leq  1.9 \cdot 10^{-1}~\mathrm{eV}$). If future cosmological data will establish this range for $m_0$ 
and the measured $0\nu\beta\beta$-decay half-lives will be within the range 
given by inequality (\ref{final}) this will be an evidence in favor of 
the Majorana mass mechanism. It is obvious that without knowledge of 
the value of the lightest neutrino mass it is impossible to determine 
in which region we are (IS or NS).
\item
 In addition to the Majorana mass mechanism, also other mechanisms
of the $0\nu\beta\beta$-decay, caused by a possible violation of the
total lepton number $L$ at a scale which is much smaller than
the standard seesaw  GUT  scale, were discussed in the literature. If
$L$ is violated at $\sim$ TeV scale a contribution of these additional
mechanisms to the matrix element of the $0\nu\beta\beta$-decay can
be comparable with the Majorana mass contribution.\footnote{In fact, the matrix element of
the $0\nu\beta\beta$-decay in the case of the Majorana neutrino mass mechanism is proportional
to $(\frac{G_{F}}{\sqrt{2}})^{2}\frac{m_{\beta\beta}}{p^{2}}$, where $p\simeq 100$ MeV
is the average momentum of the virtual neutrino. On the other side, the contribution to
the matrix element of the $0\nu\beta\beta$-decay by an
exchange of a heavy Majorana lepton with a mass $M_{\chi}\sim \Lambda$
($\Lambda$ is the scale of new physics)  is proportional to
$(\frac{G_{F}}{\sqrt{2}})^{2}(\frac{M^4_{W}}{\Lambda^{5}})$
($M_W$ is the mass of the W-boson) \cite{Vissani}. If we
assume that  $| m_{\beta\beta}|\simeq 10^{-2}$ eV we conclude that both contributions
are comparable, if $M_{\chi}\simeq 2$ TeV.}.
Thus, a significant violation of the inequalities (\ref{final}) could happen.
Examples of new mechanisms are the exchange of heavy Majorana neutralino
or gluino in SUSY models with the violation of the R-parity
(see recent papers \cite{mohap,difmech,lisi11,petcov11}), the exchange
of right-handed Majorana neutrinos with mass at the electroweak scale
(see \cite{Ibarra}), the exchange of heavy Majorana right-handed neutrinos
in $L-R$ models (see \cite{Vissani}), etc.

New mechanisms are characterized by parameters which are not connected
with neutrino oscillation parameters. In addition, the NMEs of the Majorana
mass mechanism and these possibly additional mechanisms are not connected
and are different. Thus, it requires fine tuning for such mechanisms
to contribute to the relatively narrow Majorana mass region (\ref{invhierar1})
and mimic the effect of the Majorana phase difference. It is natural to expect
that contributions to the matrix element of the $0\nu\beta\beta$-decay of the
Majorana mass mechanism and mechanisms connected with the violation of the
lepton number at TeV scale could be quite different (see, for example,
\cite{Ibarra,Senjanovic}). Moreover, such mechanisms will be checked in
LHC experiments by the search for effects of a violation of the total
lepton number (like production of the same-sign lepton pairs in $p-p$ collisions,
etc., see \cite{Senjanovic}).
\end{enumerate}

 In Fig.\ref{fig.2} we present ranges of $0\nu\beta\beta$-decay
half-lives of different nuclei in the case of inverted hierarchy
of the neutrino masses.
Three sets of NMEs are considered:\\
1) The ISM NMEs \cite{lssm} calculated for $g_A=1.25$ (see second column of
Table \ref{tab.1}). In this case $M(A,Z)= M^{\mathrm{min}}(A,Z) = M^{\mathrm{max}}(A,Z)$.\\
2) The minimal $M^{\mathrm{min}}(A,Z)$ and maximal $M^{\mathrm{max}}(A,Z)$ NMEs of
the QRPA, IBM and EDF approaches (see Table \ref{tab.1}). \\
3) The NMEs  calculated in the framework of the
renormalized QRPA and the QRPA ((R)QRPA)
with their variances obtained with Brueckner short-range correlations.
The effective weak coupling constant is assumed to be within the range
$1.0 \le  g^{eff}_A \le 1.25$ \cite{src}.
Here, $g^{eff}_A$ is the quenched axial-vector coupling constant.
We note that the significant reduction
(quenching) of the strength observed in nuclear Gamow-Teller transitions
still has no clear experimental quantification and theoretical understanding.
Usually, two possible physical origins of the quenching have been discussed in the
literature: one due to the $\Delta$-isobar admixture in the nuclear wavefunction and
another one due to the shift of the Gamow-Teller strength to higher excitation
energies induced by short-range tensor correlations. In the absence of a better
prescription, the effect of quenching is often simply evaluated by replacing the
bare value $g_A =1.25$ with an empirical, quenched value $g^{eff}_A =1$. \\

Considering the ISM NMEs, the allowed intervals for half-lives are purely
determined by the Majorana phase difference $\alpha_{12}$ with
$0 \le \sin^2\alpha_{12} \le 1$. From Fig. \ref{fig.2} we see that
the allowed ranges of $T^{0\nu}_{1/2}$ calculated with the QRPA, IBM and EDF NMEs
differ significantly from the ISM ranges.
For nuclei, where a comparison is possible, the ranges are slightly larger, however,
shifted  down to smaller half-lives. It reflects the difference in calculated NMEs between
the ISM and a group of the QRPA, IBM and EDF approaches for $g_A = 1.25$.
We notice that for many nuclei  the ranges of half-lives obtained with the
(R)QRPA results by assuming a possible quenching of $g_A$ in nuclear matter
(the third set of NMEs)  are comparable with those of the QRPA, IBM and EDF NMEs
for $g_A$=1.25. Thus, the problem of the correct description of two-nucleon
short-range correlations and of the determination of the effective weak-coupling
constant $g^{eff}_A$ is as important  as the differences incorporated in
the construction of nuclear wave functions. However, it is not clear yet how
to determine $g^{eff}_A$ reliably for a given nucleus \cite{lisi}.

In the last few years there has been a significant progress in
understanding the source of the spread of calculated NMEs. Nevertheless,
there is no consensus as yet among nuclear theorists on the
correct NME values and their uncertainties. However, the recent development in the field is
encouraging. There is good reason to believe that the uncertainty will be reduced.

We will finish with the following remarks:
\begin{enumerate}
  \item We have considered the $0\nu\beta\beta$-decay
in the case of the inverted spectrum with $m_0\lesssim 5\cdot 10^{-2}$ eV. If $m_0\gtrsim 1\cdot 10^{-1}$ eV  the neutrino mass spectrum is quasi-degenerate
\begin{equation}\label{degenerate}
m_{1}\simeq m_{2}\simeq m_{3}.
\end{equation}
The effective Majorana mass is relatively large in this case  and for both types of the neutrino mass spectrum is given by the expression
\begin{equation}\label{degenerate1}
m_{0}(\cos^{2}\theta_{13} \cos2\theta_{12}- \sin^{2}\theta_{13})\leq |m_{\beta\beta}|\leq m_{0}.
\end{equation}

The allowed region for $m_{\beta\beta} $ is presented by the region between two parallel lines in the upper part of Fig.\ref{fig.1}. It is evident  that the Majorana mass mechanism of the $0\nu\beta\beta$-decay can be checked also in the case of the quasi-degenerate spectrum if $m_{0}$ is known (from cosmological observations or from future $\beta$-decay experiments KATRIN\cite{Katrin,otten} and MARE\cite{Mare}).
\item
  In the case of normal neutrino mass spectrum with $m_0\ll \sqrt{\Delta m^{2}_{S}}\simeq 8.7\cdot 10^{-3}$ eV
 we have normal mass hierarchy\footnote{For the sum of neutrino masses we have $\Sigma m_{i}\simeq 6\cdot 10^{-2}$ eV in this case.}
 \begin{equation}\label{normal}
m_{1}\ll m_{2}\ll m_{3}.
\end{equation}
The effective Majorana mass is given in this case by the expression
\begin{eqnarray}\label{normal1}
&&|m_{\beta\beta}|\simeq \nonumber  \\
&&|\cos^{2} \theta_{13}\, \sin^{2} \theta_{12}\, \sqrt{\Delta
m^{2}_{S}}+\sin^{2} \theta_{13}\, \sqrt{\Delta m^{2}_{A}}e^{2i\alpha_{23}}| \nonumber  \\
&&\leq 3.9\cdot 10^{-3}~\mathrm{eV}.
\end{eqnarray}
Thus, in the case of the normal mass hierarchy $|m_{\beta\beta}|$
is too small in order to be probed in the $0\nu\beta\beta$-decay experiments of the next generation.
\item
The check of the  Majorana neutrino
mass mechanism which we have discussed is possible even if an observation of the
$0\nu\beta\beta$-decay is made for only one isotope.
If the half-lives of the $0\nu\beta\beta$-decay of several isotopes will
be measured and  all measured half-lives satisfy the inequality (\ref{final}), this will be additional evidence in favor of the Majorana
mass mechanism. Let us also notice that if the NME problem will be finally solved than  (independently of the character
of the neutrino mass spectrum and the value of the lightest neutrino mass) an information about the Majorana neutrino mass mechanism could be obtained
by comparing the values of the effective Majorana mass
determined from the measured half-lives of different nuclei. If it would
occur that the value of $|m_{\beta\beta}|$ does not depend on the nucleus it
would be an argument in favor of the seesaw origin of Majorana neutrino masses.

\end{enumerate}

\section{Conclusion}

From the standard seesaw mechanism, based on the assumption that the total
lepton number is violated at the  $(10^{14}-10^{15})$ GeV scale,
follows that the Majorana neutrino exchange mechanism is the only mechanism
of the neutrinoless double $\beta$-decay. In this case the half-life of
the $0\nu\beta\beta$-decay is given by the general
expression (\ref{halflife}). Thus, the test of the expression (\ref{halflife})
could give us information about the validity of the standard seesaw mechanism.
We have demonstrated that this check
could be performed if $0\nu\beta\beta$-decay will be observed in future
experiments sensitive  to the effective Majorana mass in the inverted mass hierarchy  
region, if  a present cosmological bound on the sum of the neutrino mass 
will be  improved down to 0.2 eV. For this case we
calculated the allowed ranges of the half-lives of the  $0\nu\beta\beta$-decay of
$^{48}Ca$, $^{76}Ge$ and other nuclei of experimental interest with the NMEs
obtained in the framework of all methods existing at present.
Of course, further progress in the calculation of the NMEs
is desired to obtain more precise ranges for (\ref{final}).

\acknowledgments

The work of S.B. and W.P. is supported by the Deutsche For\-schungsgemeinschaft
(Transregio 27: Neutrinos and Beyond), the Munich Cluster of Excellence
(Origin and Structure of the Universe), and the  Maier-Leibnitz-Laboratorium
(Garching). A.F. and F.\v{S}. acknowledge support of the Deutsche Forschungsgemeinschaft
within the project 436 SLK 17/298 and of the VEGA Grant agency of
the Slovak Republic under the contract No.~1/0639/09.


\end{document}